\documentstyle[12pt]{article}

\renewcommand{\theequation}{\arabic{section}.\arabic{equation}}

\renewcommand{\thesection}{\arabic{section})}

\newcommand{\be}{\begin{equation}}
     \newcommand{\ee}{\end{equation}}
     \newcommand{\ba}{\begin{eqnarray}}
     \newcommand{\ea}{\end{eqnarray}}
     \newcommand{\spa}{\hspace{0.5cm}}

     \newfont{\myfont}{msbm10 scaled\magstep1}
     \newcommand{\del}{\partial}
\newcommand{\tende}{\stackrel{|\vec x-\vec y|\rightarrow \infty}
{\longrightarrow}}

\def\b{\varphi_{_0}}

\def\A{\tilde{{\cal A}}}

\def\om{\overline \omega}

\def\rot{\vec{\nabla} \times \om}

\newcommand{\cl}{{\cal L}}

\newcommand{\fr}{\frac}

\begin{document}

      \begin{titlepage}

\leftmargin -10mm
\topmargin -1cm

\hsize 165mm

\rightline{PUPT-1403}
\rightline{DART-HEP-93/04}
\rightline{May 1993}
\vskip 10.0mm

\centerline{\LARGE\bf Mass Spectrum and Correlation Functions}
 \vskip 3pt

\centerline{\LARGE\bf of Nonabelian Quantum Magnetic Monopoles}

    \vskip 2.0cm

    \centerline{\sc E. C. Marino \footnote{On sabbatical leave from
Departamento de F\'{\i}sica, Pontif\'{\i}cia Universidade
Cat\'{o}lica, Rio de Janeiro, Brazil.}$^{,} \;$\footnote{E-mail:
marino@puhep1.princeton.edu }}

     \vskip 0.5cm

     \centerline{\it Joseph Henry Laboratories }
    \centerline{\it   Princeton University }
    \centerline{\it    Princeton, NJ 08544 }

    \vskip  0.7cm

    \centerline{\sc Rudnei O. Ramos \footnote{E-mail:
rudnei@northstar.dartmouth.edu}}

\vskip  0.5cm

     \centerline{\it Department of Physics and Astronomy}
\centerline{\it Dartmouth College}
\centerline{\it Hanover, NH 03755}

\vskip 0.7cm




\begin{abstract}

  The method of quantization of magnetic monopoles based on the order-disorder
duality existing between the monopole operator and the lagrangian fields
is applied to the description of the quantum magnetic monopoles
of `t Hooft and Polyakov in the SO(3) Georgi-Glashow model.
  The commutator of the monopole operator with the magnetic
charge is computed explicitly, indicating that indeed the quantum
monopole carries $4\pi/g$ units of magnetic charge. An explicit
expression for the asymptotic behavior of the monopole correlation function
is derived. From this, the mass of the quantum monopole
 is obtained. The tree-level result for the quantum monopole mass
is shown to satisfy the Bogomolnyi bound
($M_{\rm mon} \geq 4 \pi \frac{M}{g^2}$) and to be within the range of
values found for the energy of the classical monopole solution.

\end{abstract}

\vfill
\end{titlepage}

\leftmargin -10mm
\topmargin -8mm
\hsize 165mm
\baselineskip 7.3mm

\setcounter{page}{2}


\section{Introduction}

\spa A few years ago a general method of quantization of nonabelian
magnetic monopoles was established \cite{mon} by exploiting the
general fact that the operator which creates the topological
excitations of a certain theory must also be dual to the basic
lagrangian fields (in the sense of order-disorder duality) \cite{evora}.
This method of quantization has been applied to a variety of
systems containing topological excitations in two, three and
four dimensional spacetime \cite{kinks,vortices,vor}.

The nonabelian monopoles are topological excitations
which occur when
a nonabelian symmetry group (with compact covering) of a gauge theory
is spontaneously broken down to a U(1) symmetry. The
topological charge of the monopoles
is the abelian magnetic charge
corresponding to the unbroken U(1) \cite{thp}. As a consequence
of the fact that monopoles are topological excitations
appearing in a process of symmetry breakdown,
it can be shown \cite{mon,evora} that
for groups with a compact covering, the quantum creation operator
of magnetic monopoles is the disorder variable for the phase
transition in which the Higgs field develops a vacuum expectation value
and thereby generates a mass to the gauge fields. In a Higgs phase,
where $\langle\phi\rangle\neq 0$, we must have the vacuum expectation value of
the monopole operator ($\mu$ operator) $\langle\mu\rangle=0$.
This automatically
implies that $\mu$ creates states which are orthogonal to the
vacuum, i.e., nontrivial states \cite{evora}.
An explicit expression for the monopole operator in terms of the
basic lagrangian fields of the theory is then constructed  by imposing
that it must satisfy an order-disorder algebra with these fields.
Also a general expression for the correlation functions of these
operators is obtained as an euclidean functional integral over the
lagrangian fields by generalizing the methods first introduced by
Kadanoff and Ceva for the description of  correlation
functions of disorder variables in the Ising model \cite{kc}.

In the present work, we consider the magnetic monopoles of the
SO(3) Georgi-Glashow
model. We take the expression
obtained in \cite{mon} for the quantum operator corresponding to
the classical monopole solution
and evaluate  the long distance
behavior of its two point correlation function by using the functional
integral methods developed in \cite{mon}.
We show that this correlation function decays exponentially and
from its explicit expression the mass of the quantum
monopole is obtained. The result generated in the lowest order in
  a loop expansion is found to be
in agreement with the Bogomolnyi bound \cite{bound}, $M_{\rm mon} \geq 4 \pi
\frac{M}{g^2}$ ($M$ is the vector gauge field mass), for the
value of the classical energy of the monopole solution.

In a recent publication \cite{vor} we performed the analogous
computation in the case of  vortices in the  Abelian Higgs model
in 2+1 D. There we also have found that at the tree level the quantum
vortex mass coincides with the classical vortex solution energy, as
obtained in \cite{bound}, for example.

The quantum description of the excitations belonging to the
topologically nontrivial sectors of the theory is of fundamental
importance in order for a complete understanding of the
system to be achieved. The obtainment of monopole correlation functions,
is a basic step in the fullfilment of this goal.
It would, on the other hand, be extremely
interesting to investigate how much important are the quantum properties
of monopoles in physical processes like the eventual monopole production
in the early universe or in the catalysis of baryon decay
\cite{callan}. We envisage these processes as interesting potential fields
of application of the results of our work.

The organization of the paper is as follows. In Sec. 2 we give a
brief description of the method of quantization of monopoles
introduced in \cite{mon}. In Sec. 3 we  implement the introduction
of the external field in the functional integral describing the
monopole correlation function, which is one of the key features of
the method. In Sec. 4 the mass of the quantum monopole is obtained
 inlowest order in a loop
expansion.
 The conclusions are presented in Sec. 5. Four
Appendixes are included in order to demonstrate useful results.




\section{The Quantization of Magnetic Monopoles}
\setcounter{equation}{0}

\spa
In this section, we are going to review the method of monopole
quantization introduced in \cite{mon}. We also
evaluate the commutator of the monopole
operator with the magnetic charge and
prove its nontriviality.

Let us consider the SO(3) Georgi-Glashow model, given by
\be
\cl = -\fr{1}{4} G_{\mu\nu}^a G^{\mu\nu a} +\fr{1}{2}(D_\mu
\phi)^a (D^\mu \phi)^a - \fr{1}{2} m^2 \phi^a \phi^a
-\fr{\lambda}{4} (\phi^a \phi^a)^2 \: .
\label{Lagradjoint}
\ee
\noindent
Throughout this work, we are going to use the adjoint representation
where the generators of SO(3) are $(T^a)^{bc}=i \epsilon^{abc}$
($a,b,c=1,2,3$),  $[T^a,T^b]=i \epsilon^{abc} T^c,\ \
{\rm tr} T^aT^b=2\delta^{ab}$. Then,
the Higgs field belongs to an isospin triplet $\phi^a$ ($a=1,2,3$) and
$$
G_{\mu\nu}^a= \del_\mu W_\nu^a -\del_\nu W_\mu^a + g \epsilon
^{abc} W_\mu^b W_\nu^c \: ,
$$
\be
(D_\mu \phi)^a =\del \phi^a + g \epsilon^{abc} W_\mu^b \phi^c \:,
\label{2.2}
\ee
\noindent
where g is the charge coupling constant. The model can exist
in two
phases according to whether $m^2>0$ or $m^2<0$. The first one is the
symmetric phase and the second one is the ``broken'' phase where
the Higgs field acquires a vacuum expectation value $|\langle\phi^a\rangle |=
\varphi_0=(|m^2|/\lambda)^{1/2}$. The theory possesses an
identically conserved topological current which can be written as
\be
J^\mu = \fr{1}{2} \epsilon^{\mu\nu\alpha\beta}\del_\nu \left[G^a_{
\alpha\beta}\fr{\phi^a}{|\phi^a|}\right] \: .
\label{2.3}
\ee
\noindent
By introducing the electromagnetic field
\be
F^{\mu\nu} = G^a_{\mu\nu}\fr{\phi^a}{|\phi|}= \fr{1}{2} {\rm tr}
G_{\mu\nu}\fr{\phi}{|\phi^a|} \: ,
\label{2.4}
\ee
\noindent
where $G_{\mu\nu}=G^a_{\mu\nu}T^a$, $W_\mu=W_\mu^aT^a$
and $\phi=\phi^a T^a$
we immediately realize that the topological charge density
$J^0=\fr{1}{2}\epsilon^{ijk}\del_i F_{jk} =\vec\nabla\cdot\vec B$
is precisely the magnetic charge. Observe that the electromagnetic
field $F_{\mu\nu}$ is gauge invariant: under $g(\alpha)=\exp[i
\alpha^a T^a]$, we have \break $G_{\mu\nu}\rightarrow gG_{\mu\nu}g^{-1}$
and $\phi\rightarrow g\phi g^{-1}$.

It was shown by `t Hooft and Polyakov \cite{thp} that in the broken
symmetry phase the theory admits classical finite energy solutions
carrying a magnetic charge $4\pi /g$. The classical mass (energy) of
the magnetic monopole solution was computed to be \cite{thp,prasad,zachos}
\break $M^{\rm cl}_{\rm mon}=
f(\frac{\lambda}{g^2}) 4 \pi \frac{M}{g^2}$ where the pure
number prefactor, $f(\frac{\lambda}{g^2})$, depends
on the  coupling constants ratio
$\lambda /g^2$ ($f(0) = 1$ \cite{prasad} and $f(\infty) = 1.787$
\cite{zachos}).
The monopole solution has the asymptotic behavior given by
a nontrivial gauge transformation out of the vacuum $\phi^a_v =\varphi_0
\delta^{a3}$, $ W^a_{i,v}=0$:
\be
\phi^a(x) \stackrel{|\vec x| \rightarrow\infty}{\sim}\varphi_0
\fr{x^a}{|\vec x|} = \varphi_0 g(\omega)^{ab}\delta^{b3} \: ,
\label{2.5}
\ee
\[
W_i^a(x)
\stackrel{|\vec x| \rightarrow\infty}{\sim} -\fr{1}{g}
\epsilon^{iab}\fr{x^b}{|\vec x|^2}
\]
\noindent
or
\be
W_i=W^a_i T^a = -\fr{i}{g} \del_i g(\omega) g^{-1}(\omega) \: ,
\label{2.6}
\ee
\noindent
where $g(\omega)$ is an SO(3) matrix with parameters given by
\cite{thp}
$$
\omega^a= \omega\:\; \left(0, \fr{\tan \fr{\theta}{2}}{\sqrt{\sin^2\phi +
\tan^2
\fr{\theta}{2}}}, \fr{\sin\phi}{\sqrt{\sin^2\phi +
\tan^2\fr{\theta}{2}}}\right) \: ,
$$
\be
\omega= 2 \arccos\left(\cos\phi \cos \fr{\theta}{2}\right) \: .
\label{2.7}
\ee

The electromagnetic field associated with the asymptotic
configuration, (2.5) and (2.6), is
\be
F^{0i}_{\rm mon}=0\ \ \ ;
\ \ \  F^{ij}_{\rm mon}=\fr{1}{g} \epsilon^{ijk} \fr{x^k}{|\vec x|^3} \: .
\label{2.8}
\ee
\noindent
Since for the vacuum $F^{\mu\nu} =0$, we see that under the
transformation $g(\omega)$,
\be
  F^{\mu\nu} \stackrel{g(\omega)}
{\longrightarrow} F^{\mu\nu} + F^{\mu\nu}_{\rm mon} \: .
\label{2.9}
\ee

The dual algebra which the magnetic monopole operator must
satisfy is related to the above asymptotic solution \cite{mon}

\be
\mu(x,S) \phi^a(y)= \left\{
\begin{array}{ll}
g(\omega)^{ab}\phi^b(y) \mu(x,S)\:\:,
&\:\: \vec y \in V_x(S)\\
\phi^a(y) \mu(x,S)\:\:, &\:\:\vec y \not\in V_x(S)
\end{array}
\right.
\label{2.10}
\ee
\be
\mu(x,S)W_i(y)= \left\{
\begin{array}{ll}
\left[ g(\omega) W_i(y) g^{-1}(\omega)
-\fr{i}{g} \del_i g(\omega) g^{-1}(\omega) \right] \mu(x,S) \:\:,
&\vec y  \in V_x(S)\\
W_i(y) \mu(x,S) \:\:,& \vec y \not\in V_x(S) \:.
\end{array}
\right.
\label{2.11}
\ee
\noindent
In the above expression $S$ is a sphere of radius $\rho$,
centered on $\vec x$ and $V_x(S)$ is the volume $\Re^3 -T(S)$
where $T(S)$ is the spherical volume bounded by $S$ (Figure 1).
Observe that as a consequence of the above algebra, $\mu$ is
in principle a nonlocal operator depending on $S$. A local
$\mu$ can be obtained in the limit when $\rho \rightarrow 0$.

An operator realization for $\mu$
can be obtained by using the external
field \cite{mon}
\be
\A_\mu(z;x) \equiv \A_\mu^a(z;x)T^a = -\fr{1}{g} \int _{V_x(S)} d^3\xi_\mu
\omega^a(\xi - x) \delta^4 (z- \xi)|_{\xi^0 =x^0} T^a \:.
\label{external field}
\ee

\noindent
In terms of this, we can write the monopole operator as \cite{mon}
$$
\mu(x;S) = \exp \left\{\fr{-i}{2} \int d^4z {\rm tr}
[D_\mu G^{\mu\nu} \A_\nu(z;x)]
\right\}\:,
$$
\be
\mu(x;S) = \exp \left\{- i \int d^4z [D_\mu G^{\mu\nu}]^a \A_\nu^a(z;x)
\right\}\:.
\label{2.13}
\ee

\noindent
Also, using the Yang-Mills equation
\be
 [D_\mu G^{\mu\nu}]^a = g j^{\nu a} \equiv g \epsilon^{abc} [D^\nu \phi]^b
\phi^c \:,
\label{2.14}
\ee

\noindent
we can write the $\mu$ operator as
\be
\mu(x;S) = \exp \left\{ -i g \int d^4z j^{\mu a} \A_\mu^a(z;x)\right\}\:.
\label{monopole operator}
\ee

\noindent
As is shown in \cite{mon} these expressions for $\mu$ satisfy the
dual algebra (2.10-2.11).

We can also compute the commutator of the monopole operator
with the magnetic charge. Let us consider the local case in which the
radius of the sphere $S$ goes to zero. Using \break (2.10-2.11) or the more
convenient form (for $\rho\rightarrow 0$)
$$
[\mu(x),W_i(y)] = \left(\tilde{W}_i(y) -W_i(y)\right)\mu(x)\:,
$$
\be
[\mu(x),\phi^a(y)]= \left(\tilde{\phi}^a(y) - \phi(y)\right) \mu(x)\:,
\label{2.16}
\ee

\noindent
it is easy to show that
\be
[\mu(x),G_{ij}^a(y)]= \left(\tilde{G}^a_{ij}(y) -G^a_{ij}(y) \right)\mu(x) \:.
\label{2.17}
\ee

\noindent
In the above equations,
$\tilde{{\cal O}}$ is the $g(\omega)$-transform of ${\cal O}$.
Using (2.16-2.17) it is straightforward to show that
\be
[\mu(x),F_{\mu\nu}(y)] = \left(\tilde{F}_{\mu\nu}(y) - F_{\mu\nu}(y)
\right) \mu(x)  \:,
\label{2.18}
\ee

\noindent
or
\be
[\mu(x),F_{\mu\nu}(y)] = F^{\rm mon}_{\mu\nu}\mu(x)\:,
\label{2.19}
\ee

\noindent
where we used (2.9). The commutator of $\mu$ with the magnetic
charge (topological charge) density is now immediately seen to be
\be
\left[ J^0(y),\mu(x) \right]= - \fr{1}{2} \epsilon^{ijk}\del^{(y)}_i
 F_{jk}^{\rm mon}(y)
\mu(x) =-\fr{1}{g} \nabla^2_{(y)}\left[\fr{1}{|\vec x -\vec y|}\right]\mu(x)=
\fr{4\pi}{g} \delta^3(\vec x-\vec y)\mu(x)\:.
\label{2.20}
\ee

\noindent
This result shows explicitly that the operator $\mu$ does indeed carry
$4\pi/g$ units of magnetic charge.

In ref. \cite{mon} a gauge equivalent form of the operator
$\mu$ was used. Consider the set of parameters of an SO(3)
transformation given by
\be
   \overline{\omega}^a = \theta \; (- \sin \phi, \cos \phi, 0)\: .
\label{omega}
\ee

\noindent
$g(\overline{\omega})$ shares with $g(\omega)$ the property
(2.5), that is,
\be
g(\overline{\omega})^{ab} \delta^{b3} = \fr{x^a}{|\vec x|}\:.
\label{2.22}
\ee

\noindent
On the other hand, if we consider the configuration $\overline{W}_i=
-\fr{i}{g} \del_i g(\overline\omega) g^{-1}(\overline\omega)$, it
follows that it must be gauge equivalent to $W_i$, Eq. (2.6),
because both are gauge transforms of the vacuum ($W_{i,v}=0$)
\footnote{One of the authors (E.C.M.) is grateful to
A. di Giacomo for calling his attention to the fact
that the field configuration $W_\mu$ obtained by a gauge transformation
of the vacuum with $g(\overline\omega)$ is not identical
to the one obtained with $g(\omega)$.}. This
means that $\overline{W}_i = h W_i h^{-1} -
\fr{i}{g} \del_i h h^{-1}$, where $g(\overline
\omega) = h g(\omega)$ and $h^{ab}\fr{x^b}{|\vec x|}=\fr{x^a}{|\vec
x|}$. Another consequence is that the field intensity tensor
configuration associated with $\overline{W}_i$ must be  related
to the one
 associated with $W_i$ as $\overline{G}_{\mu\nu}=
h G_{\mu\nu} h^{-1}$. It follows that the electromagnetic field
obtained out of the vacuum ($F^{\mu\nu}=0$) by $g(\overline\omega)$
is
\be
F_{\mu\nu}=0 \stackrel{ g(\overline\omega)} \longrightarrow
\fr{1}{2} {\rm tr} h G_{\mu\nu} h^{-1} \hat{\phi}_r=
\fr{1}{2} {\rm tr}  G_{\mu\nu} \hat{\phi}_r= F^{\rm mon}_{\mu\nu}  \:,
\label{2.23}
\ee

\noindent
where $F^{\mu\nu}_{\rm mon}$ is given by (2.8),
${\hat \phi}^a_r\equiv x^a/|\vec x|$ and we have used the cyclic
property of the trace as well as the fact that $\phi^a_r$ is
invariant under $h$ (or $h^{-1}$). As a consequence of (2.23) and (2.18-2.19)
it follows that the operator $\mu$ constructed with $\overline
\omega$ bears the same magnetic charge as the one constructed
with $\omega$. The operator $\mu(\overline\omega)$ is gauge
equivalent to $\mu(\omega)$. Troughout this work, we are going to
use  the operator $\mu$ expressed in terms of
$\overline\omega$ because of its more convenient form.

The monopole operator $\mu$ is in principle nonlocal because
it is defined on the volume (tridimensional hypersurface in
four dimensional space)
$V_x(S)$. A local operator, however,
can be obtained by the introduction of an appropriate renormalization
factor \cite{mon} and by taking the limit when $\rho$, the radius of
$S$ goes to zero.
 When computing correlation functions  of $\mu$,
this is naturally done within the euclidean functional integral framework,
by treating $\mu$ as a disorder variable and imposing hypersurface
invariance
on the expression for the correlation functions \cite{mon}. The
hypersurface dependent renormalization counterterms appear  then
as self interactions of the external field $\A_\mu(z;x)$. Here we
reproduce the final results for the hypersurface independent
$\mu$ correlation functions and refer the reader to \cite{mon}
for further details.
The $\mu$ two point correlation function is given by (Euclidean space)
\ba
\langle\mu(x)\mu^\dagger(y)\rangle &=&\lim_{\rho \rightarrow 0}
Z^{-1}[0]
\int DW_\mu D\phi D\eta D\overline\eta
\exp \left\{- \int d^4z \left[-\fr{1}{8} {\rm tr}
\left\{ G_{\mu\nu}[W_\mu +
 \right. \right. \right. \nonumber \\
&+& \left. \left. \left.
\A_\mu(z;x,y)]\right\}^2 +
 \fr{1}{2} (D_\mu \phi)^a (D^\mu \phi)^a +
V(\phi) +\cl_{\rm GF}+
\cl_{\rm gh}\right] \right\} \:.
\label{2.24}
\ea

\noindent
In this expression $\A_\mu(z;x,y) = \A_\mu(z;x) - \A_\mu(z;y)$, where
$\A_\mu(z;x)$ is given by (2.12). $\cl_{\rm GF}$ and $\cl_{\rm gh}$ are the
gauge fixing and ghost terms, respectively and $\eta$ and
$\overline\eta$ are the ghost fields. As is shown in \cite{mon},
the above expression is hypersurface invariant and therefore local
as a consequence of gauge invariance (or BRST invariance),
because we can change the
hypersurface $V_x(S)$ by means of a gauge transformation.

By making the change of variable $W_\mu \rightarrow
W_\mu - \A_\mu(z;x,y)$ in the above functional integral we immediately
obtain the equivalent form for the $\mu$ two point function:
\ba
\langle\mu(x)\mu^\dagger(y)\rangle &=& \lim_{\rho \rightarrow 0} Z^{-1}[0]
\int D W_\mu D\phi D\eta D\overline\eta
\exp \left\{- \int d^4z \left[-\fr{1}{8} {\rm tr}\{ G_{\mu\nu}\}^2
+ \right. \right. \nonumber \\
&+& \left. \left. \fr{1}{2} (\tilde{{\cal D}}_\mu \phi)^a
(\tilde{{\cal D}}^\mu \phi)^a + V(\phi) +\cl_{GF}(W_{\mu}^a \to
W_{\mu}^a - \A_{\mu}^a) + \right. \right. \nonumber \\
&+& \left. \left. \cl_{gh}(W_{\mu}^a \to W_{\mu}^a
- \A_{\mu}^a)\right] \right\} \:,
\label{correlation2}
\ea

\noindent
where $\tilde{{\cal D}}_\mu ={\bf 1} \del_\mu - ig\left[ W_\mu -
\A_\mu(z;x,y)\right]$.
By dropping the renormalization self interaction terms of the
external field
from (2.24) and (2.25) we can immediately recognize the expressions
(2.13)  and (2.15), respectively, for the monopole operator $\mu$.
These expressions are going to be our starting point for the
obtainment of the long distance behavior of the monopole
correlation function and mass. The great advantage of them is that
their computation reduces to a standard computation in a field
theory in the presence of the external field $\A_\mu$.

\section{Introducing the External Field $\A_{\mu}^{a}(z)$ in
the Broken and Symmetric Phases}

\setcounter{equation}{0}

\spa
In this section we study the introduction of the external field
$\A_{\mu}^{a}(z)$, used in the description of the magnetic monopole
correlation function, in both phases of the $SO(3)$ Georgi-Glashow model:
(a) the symmetric phase, with the mass parameter
in (\ref{Lagradjoint}) $m^2 > 0$ (with $\langle \phi \rangle= 0$, for
the vacuum expectation value of the Higgs field); and (b) the broken
phase, with mass parameter $m^{2} < 0$ ($\langle \phi \rangle \neq 0$).

\subsection{Broken and symmetric phases}

\spa
In the symmetric phase the Lagrangian density, ${\cal L}^S$, is just given by
(\ref{Lagradjoint}) and there is no need to make any shift in
the Higgs fields around the vacuum.
In the symmetric phase we have only to add to (\ref{Lagradjoint}) a
gauge-fixing term ${\cal L}_{GF}$ along with the corresponding ghost
term ${\cal L}_{\rm gh}$. For the gauge-fixing term, we may choose a
Lorentz-type gauge. Then, we add to (\ref{Lagradjoint}), in the
symmetric phase, the terms

\[
{\cal L}_{GF}^{S} = - \frac{\xi}{2} \left(\partial_{\mu} W_{a}^{\mu}
\right)^{2} \: ,
\]
\begin{equation}
\label{gaugeghost}
\end{equation}
\[
{\cal L}_{\rm gh}^{S} = - \bar{\eta}^{a} \left( \delta^{ab} \Box -
g \varepsilon^{abc} \partial^{\mu} W_{\mu}^{c} -
g \varepsilon^{abc} W_{\mu}^{c} \partial^{\mu} \right) \eta^{b} \: ,
\]

\noindent
where $\eta^{a}$ are ghost fields and $\xi$ is the gauge parameter.

In the broken phase, $m^{2} < 0$ in (\ref{Lagradjoint}), the
potential $V(\varphi^a \varphi^a) = \frac{m^2}{2} \varphi^a \varphi^a
+ \frac{\lambda}{4}(\varphi^a \varphi^a)^2 $ has
a minimum at
$(\varphi^a \varphi^a) = \b^2 $, with
$\b^2 = \frac{|m^2|}{\lambda}$. Choosing
the vacuum pointing along the third isospin axis, that is,
$\varphi^a = \b \delta^{a \: 3}$ and shifting the fields around
this value,  we see
that the physical fields will be given by
$(\phi_{1},\phi_{2},\phi_{3}) \rightarrow (\phi_{1},\phi_{2},\chi)$,
with $\chi = \phi_{3} - \b$.
The Lagrangian density in the broken phase is then given,
after shifting, by

\begin{eqnarray}
{\cal L}^{B} &=& - \frac{1}{4} G_{\: \: \mu \nu}^{a} G^{a \: \mu \nu} +
\frac{1}{2} \left[ \left(\partial_{\mu} \phi_{1} \right)^2 +
\left(\partial_{\mu} \phi_{2} \right)^2 \right] + \frac{1}{2}
\left[ \left( \partial_{\mu} \chi \right)^2 - m_{\chi}^{2} \chi^{2}
\right] + \nonumber \\
&+& \frac{M^2}{2} \left[ (W_{1}^{2})^2 + (W_{2}^{2})^2 \right] +
M \left[ W_{2}^{\mu}(\partial_{\mu} \phi_{1}) - W_{1}^{\mu}
(\partial_{\mu} \phi_{2}) \right] + g \left[ W_{1}^{\mu} (\phi_{2}
\partial_{\mu} \chi - \chi \partial_{\mu} \phi_{2}) + \right. \nonumber \\
&+& \left. W_{2}^{\mu} ( \chi \partial_{\mu} \phi_{1} - \phi_{1} \partial_{\mu}
\chi)
+ W_{3}^{\mu} ( \phi_{1} \partial_{\mu} \phi_{2} - \phi_{2}
\partial_{\mu} \phi_{1}) \right] + \nonumber \\
&+& \frac{g^2}{2} \left[ (W_{1}^{\mu})^{2} (\phi^{2} + \chi^{2} + 2 \b \chi) +
(W_{2}^{\mu})^2 ( \phi_{1}^{2} + \chi^2 + 2 \b \chi) +
(W_{3}^{\mu})^2 ( \phi_{1}^{2} + \phi_{2}^{2}) \right] + \nonumber \\
&-& g^2 \left[ (W_{1}^{\mu} W_{2}^{\mu}) \phi_{1} \phi_{2} +
(W_{1}^{\mu} W_{3}^{\mu}) (\phi_{1} \chi + \b \phi_{1}) +
(W_{2}^{\mu} W_{3}^{\mu}) (\phi_{2} \chi + \b \phi_{2}) \right] + \nonumber \\
&-& \frac{\lambda}{4} \left( \phi_{1}^{2} + \phi_{2}^{2} + (\b + \chi)^2
\right)^2  \: ,
\label{Lagrbroken}
\end{eqnarray}

\noindent
where, in the expression above, $m_{\chi}^{2} = 2 \lambda \b^2$ and
$M^2 = g^2 \b^2$. The fields $\phi_{1}$, $\phi_{2}$ and $\chi$ have
zero vacuum expectation value. In the broken phase we choose a
$R_{\xi}$ gauge (or 't Hooft gauge) \cite{zuber}, where the quadratic
mixed terms involving $(W_{\mu} \phi)$ in (\ref{Lagrbroken})
disappear. In the broken phase we have then for ${\cal L}_{GF}$ and
${\cal L}_{\rm gh}$ (adjoint representation) the expressions

\[
{\cal L}_{GF}^{B} = - \frac{\xi}{2} \left[ \partial^{\mu} W_{\mu}^{a} +
\frac{M}{\xi} \varepsilon^{a b 3} \phi_{b} \right]^{2} \: ,
\]
\begin{equation}
\label{gaugeghostbroken}
\end{equation}
\[
{\cal L}_{\rm gh}^{B} = - \bar{\eta}^{a} \left[ \delta^{ab} \Box
- g \varepsilon^{abc} \partial^{\mu} W_{\mu}^{c} -
g \varepsilon^{abc} W_{\mu}^{c} \partial^{\mu} +
\left(\frac{M}{\xi} \phi_{3} + \frac{M^2}{\xi} \right) \left( \delta^{ab} -
\delta^{a3} \delta^{b3} \right)  \right] \eta^{b} \: .
\]

{}From Eqs. (\ref{Lagradjoint}) and (\ref{gaugeghost}), the effective
Lagrangian density in the symmetric phase is then given by
${\cal L}_{\rm eff}^{S} = {\cal L}^{S} + {\cal L}_{GF}^{S} +
{\cal L}_{\rm gh}^{S}$, while in the broken phase, from Eqs. (\ref{Lagrbroken})
and (\ref{gaugeghostbroken}), the effective Lagrangian density is
${\cal L}_{\rm eff}^{B} = {\cal L}^{B} + {\cal L}_{GF}^{B} +
{\cal L}_{\rm gh}^{B}$. From the quadratic terms in ${\cal L}_{\rm eff}^{S}$
and ${\cal L}_{\rm eff}^{B}$, we obtain the respective propagators for
the fields. In Euclidean space these propagators are

\begin{eqnarray}
& \Delta_{(i)}(x) & = \int \frac{d^4 k}{(2 \pi)^4} e^{i k.x} \frac{1}{
k^2 + m_{i}^{2}} \:, \:\:\:\: i=1,2,3  \: ,
\nonumber \\
& D_{(1)}^{\mu \nu}(x) & = D_{(2)}^{\mu \nu}(x) = \int \frac{d^4 k}{(2 \pi)^4}
e^{i k.x} \frac{1}{k^2 + M^2} \left( \delta^{\mu \nu} -
\frac{(\xi - 1) k^{\mu} k^{\nu}}{\xi k^2 + M^2} \right) \: ,
\nonumber \\
& D_{(3)}^{\mu \nu}(x) & = \int \frac{d^4 k}{(2 \pi)^4} e^{i k.x} \frac{1}{k^2}
\left( \delta^{\mu \nu} - \frac{(\xi -1) k^{\mu} k^{\nu}}{\xi k^2}\right) \: ,
\nonumber \\
& \Delta_{\rm gh}^{(i)}(x) & =  \int \frac{d^4 k}{(2 \pi)^4} e^{i k.x}
\frac{1}{k^2 + m_{\rm gh_{(\it i)}}^{2}} \:, \:\:\: i=1,2,3 \: ,
\label{propagators}
\end{eqnarray}

\noindent
where $\Delta_{(i)}(x)$ are the propagators for the Higgs-field components,
$\phi_{1}$, $\phi_{2}$ and $\phi_{3}$ ($\chi$ in the broken phase),
$D_{(a)}^{\mu \nu}(x)$ are the propagators for the gauge fields $W_{\mu}^{a}$
and $\Delta_{\rm gh}^{(i)}(x)$ are the propagators for ghost-field components.
In the symmetric phase we have $M=0$ and \break $m_{i}^{2} = m^2$
and
$m_{\rm gh_{(\it i)}}^{2} = 0$ ($i = 1,2,3$). In the broken phase we have
$m_{1}^{2} = m_{2}^{2} = \frac{M^2}{\xi}$, $m_{3}^{2} = m_{\chi}^{2}$ and
$m_{\rm gh_{(1)}}^{2} = m_{\rm gh_{(2)}}^{2} = \frac{M^2}{\xi}$ and
$m_{\rm gh_{(3)}}^{2} = 0$.

\subsection{Introducing the external field in ${\cal L}_{\rm eff} =
{\cal L} + {\cal L}_{GF} + {\cal L}_{\rm gh}$}

\spa
As shown in Sec. 2, the two-point monopole correlation
function $\langle \mu \mu^{\dagger} \rangle$ is given by Eq. (2.24), or,
through a change of functional integration variable,
$W_{\mu} \to  W_{\mu} - \A_{\mu}$,
by the equivalent form given by Eq. (\ref{correlation2}).
We are going to use
Eq. (\ref{correlation2}) for the evaluation of the two-point magnetic
monopole correlation function. Let us write the exponent in
(\ref{correlation2}) as $S_{\rm eff} = \int d^4 z \left[
{\cal L}_{\rm eff}^{\rm Eucl} + \tilde{{\cal L}}_{\rm eff}
(\A_{\mu}) \right]$, where $\tilde{{\cal L}}_{\rm eff}
(\A_{\mu})$ contains all the dependence on the external
field $\A_{\mu}(z;x,y)$.

In the symmetric phase, from Eqs. (\ref{Lagradjoint}), (\ref{gaugeghost})
and (\ref{correlation2}),
we obtain that (in Euclidean space)

\begin{eqnarray}
\tilde{{\cal L}}_{\rm eff}^{S}(\A_{\mu}) &=&
- g \varepsilon^{abc} \A_{\mu}^{b} \phi^{c} \partial_{\mu} \phi^{a} +
\frac{1}{2} g^2 \A_{\mu}^{a} \A_{\mu}^{a} \phi^b \phi^b
- g^2 \A_{\mu}^{a} W_{\mu}^{a} \phi^b \phi^b + \nonumber \\
&-& \frac{1}{2} g^2 \A_{\mu}^{a} \A_{\mu}^{b} \phi^a \phi^b +
g^2 \A_{\mu}^{a} W_{\mu}^{b} \phi^a \phi^b +
\frac{\xi}{2} \left[ \left(\partial_{\mu} \A_{a}^{\mu} \right)^2
- 2 \partial_{\mu} \A_{a}^{\mu} \partial_{\nu} W_{a}^{\nu} \right]
+ \nonumber \\
&+&  \bar{\eta}^{a} \left( g \varepsilon^{abc} \partial^{\mu}
\A_{\mu}^{c} + g \varepsilon^{abc} \A_{\mu} \partial^{\mu}
\right) \eta^{b} \: ,
\label{externalsym}
\end{eqnarray}

\noindent
while in the broken phase, from Eqs. (\ref{Lagrbroken}) and
(\ref{gaugeghostbroken}), we obtain (in Euclidean space)

\begin{eqnarray}
\lefteqn{\tilde{{\cal L}}_{\rm eff}^{B}(\A_{\mu}) =
\frac{M^2}{2} \left[ \left( \A_{1}^{\mu} \right)^2 - 2
\A_{1}^{\mu} W_{1}^{\mu} + \left(\A_{2}^{\mu} \right)^2 -
2 \A_{2}^{\mu} W_{2}^{\mu} \right]
- g \left[ \A_{1}^{\mu} \left( \phi_{2} \partial_{\mu} \chi
- \chi \partial_{\mu} \phi_{2} \right) + \right.} \nonumber \\
& & + \left. \A_{2}^{\mu} \left( \chi \partial_{\mu} \phi_{1} -
\phi_{1} \partial_{\mu} \chi \right) + \A_{3}^{\mu}
\left( \phi_{1} \partial_{\mu} \phi_{2} - \phi_{2} \partial_{\mu}
\phi_{1} \right) \right] + \nonumber \\
& &
+ \frac{g^2}{2} \left[ \left( \A_{1}^{\mu} \A_{1}^{\mu} -
2 \A_{1}^{\mu}   W_{1}^{\mu} \right) \left( \phi_{2}^{2} +
\chi^2 + 2 \b \chi \right) + \right. \nonumber \\
& & + \left. \left(\A_{2}^{\mu} \A_{2}^{\mu} -
2 \A_{2}^{\mu} W_{2}^{\mu} \right) \left( \phi_{1}^{2} + \chi^2 +
2 \b \chi \right) + \left( \A_{3}^{\mu} \A_{3}^{\mu} -
2 \A_{3}^{\mu} W_{3}^{\mu} \right) \left( \phi_{1}^{2} + \phi_{2}^{2}
\right) + \right. \nonumber \\
& & - \left. 2 \left( \A_{1}^{\mu} \A_{2}^{\mu} -
\A_{1}^{\mu} W_{2}^{\mu} - \A_{2}^{\mu} W_{1}^{\mu} \right)
\phi_1 \phi_2 -
2 \left( \A_{1}^{\mu} \A_{3}^{\mu} -
\A_{1}^{\mu} W_{3}^{\mu} - \A_{3}^{\mu} W_{1}^{\mu} \right)
\left( \phi_{1} \chi + \b \phi_{1} \right) + \right. \nonumber \\
& & - \left. 2 \left( \A_{2}^{\mu} \A_{3}^{\mu} -
\A_{2}^{\mu} W_{3}^{\mu} - \A_{3}^{\mu} W_{2}^{\mu} \right)
\left( \phi_{2} \chi + \b \phi_{2} \right) \right]+
\frac{\xi}{2} \left[ \left(\partial_{\mu} \A_{a}^{\mu} \right)^2
- 2 \partial_{\mu} \A_{a}^{\mu} \partial_{\nu} W_{a}^{\nu}
\right] + \nonumber \\
& & + \:  \bar{\eta}^{a} \left( g \varepsilon^{abc} \partial^{\mu}
\A_{\mu}^{c} + g \varepsilon^{abc} \A_{\mu}^{c} \partial^{\mu}
\right) \eta^b \: .
\label{externalbroken}
\end{eqnarray}

{}From Eq. (\ref{external field}) and the expression for the function
$\om^a \equiv \om^{a}(\vec{z}-\vec{r})$, Eq. (\ref{omega}),
since $\om^{3}(\vec{z}-\vec{r})=0$,
we have $\A_{3}^{\mu}(z;x,y) =0$ in Eqs. (\ref{externalsym}) and
(\ref{externalbroken}). From $\tilde{\cal L}_{\rm eff}^{S}(\A_{\mu})$
and $\tilde{\cal L}_{\rm eff}^{B}(\A_{\mu})$,
we may extract the Feynman rules involving the external field. For example,
at the tree-level order the relevant vertices are shown in Fig. 2.


\section{The Monopole Correlation Function
and Mass}

\setcounter{equation}{0}

\subsection{The contribution at lowest order}


\spa
Let us consider in this section the evaluation of the
magnetic monopole two-point correlation function. Our starting point
will be expression (2.25), as introduced in the last section.
{}From (2.25),
we can immediately see that
\be
\langle \mu(x)\mu^\dagger(y)\rangle =\exp \left\{ \Lambda (x-y)\right\}\: ,
\label{4.1}
\ee

\noindent
where $\Lambda(x-y)$ is the sum of all Feynman graphs with the
external field $\A^a_\mu(z;x,y)$ in the external legs.

In order to evaluate $\Lambda(x-y)$ we are going to use a loop
expansion. In this work, we obtain the 0-loop result.
 In a forthcoming publication, we intend to consider the
one-loop correction to this result \cite{1l}.

 We will be interested in the
large distance behavior of (4.1), namely, when $|\vec x-\vec y|
\rightarrow \infty$. As we show in Appendix A, only two legs
graphs contribute to $\Lambda$ in this limit.
At 0-loop level, the two legs graphs containing the external field
$\A^a_\mu$ are depicted in Fig. 3. Observe that the three last graphs
of Fig. 3 only occur in the broken symmetry
phase where the Higgs field possesses a nonzero vacuum expectation
value $\b$ and the gauge field acquires a mass
$M = g\b$ through the Higgs mechanism. In the symmetric phase, the
contribution to $\Lambda(x-y)$ is given only by the first two
graphs of Fig. 3.
The sum of these graphs, in the symmetric phase, is easily seen to vanish by
using the gauge field propagators given in Eq. (\ref{propagators}). This result
immediately leads us to the conclusion that in the symmetric
phase, where $\b=0$ and the additional graphs are absent,
$\Lambda(x-y) \rightarrow 0$ at large distances, implying that
\be
\langle\mu(x)\mu^\dagger(y)\rangle_S \tende 1 \: .
\label{4.2}
\ee

\noindent
This result implies that $\langle\mu\rangle\neq 0$ in the symmetric phase
expressing the fact that the $\mu$ operator does not create
states orthogonal to the vacuum  not being, therefore, a truly
monopole interpolating operator in this phase. This is an
expected result in a phase where no classical monopole solution
exists.

In the broken phase ($m^2 < 0$), from the graphs in Fig. 3 and the vertices
in Fig. 2, we can write explicitly the asymptotic contribution to
$\Lambda(x-y)$ as

\ba
\Lambda^{\rm asy}(x-y) &=& \sum_{a = 1}^{2} \int d^4 z d^4 z'
\A_{\mu}^a (z;x,y) \left[ - \frac{\xi}{2} \partial_{\mu}
\partial_{\nu} ' \delta^4 (z - z') + \right. \nonumber \\
&+& \left. \frac{\xi^2}{2} \partial_{\mu} \partial_{\alpha}
\partial_{\beta} ' \partial_{\nu} ' D_{(a)}^{\alpha \beta}(z - z')
- \frac{M^2}{2} \delta^4 (z - z') -
\xi M^2 \partial_{\mu} \partial_{\alpha} D_{(a)}^{\alpha \nu}
(z - z') + \right. \nonumber \\
&+& \left. \frac{M^4}{2} D_{(a)}^{\mu \nu}(z - z') \right]
\A_{\nu}^a (z';x,y) \: ,
\label{Lambda tree}
\ea

\noindent
where $D_{(a)}^{\mu \nu}(z)$ is the Euclidean gauge propagator given
in (\ref{propagators}). Substituting $D_{(a)}^{\mu \nu}(z)$
in (\ref{Lambda tree}), we get the result

\be
\Lambda^{\rm asy}(x-y)=-\fr{M^2}{2}
\int d^4z d^4z' \A^a_\mu(z;x,y)\left[
-\Box \delta^{\mu\nu}+\del^\mu \del^\nu \right] F(z-z')
\A^a_\nu(z';x,y) \: ,
\label{4.4}
\ee

\noindent
where

\be
F(z-z')= \int \fr{d^4k}{(2\pi)^4} \fr{e^{ik\cdot (z-z')}}{k^2 + M^2} \: .
\label{4.5}
\ee

\noindent
Observe that all the dependence on the gauge parameter $\xi$ cancels and
therefore (\ref{4.4}) is completely gauge independent.

Let us observe now that in the adjoint representation, in which
we are working, the internal indexes of $\overline\omega^a$
(remember we are using the external field Eq. (2.12), with
$\overline\omega$ given by Eq. (2.21)) or
$\A_\mu^a$ are of the same type as the spatial indexes. Since the
spacetime index $\mu$ of $\A_\mu$ is always temporal because of
our choice of the hypersurface $V_x(S)$ and since the indexes of
$\overline\omega^a$ are always spatial, we can write (4.4) as

\ba
\Lambda^{\rm asy}(x-y) &=&
-\fr{M^2}{2} \int d^4z d^4z' \A^\alpha_ \mu(z;x,y)
\left[-\Box
(\delta^{\mu\nu} \delta^{\alpha\beta} -
\delta^{\mu\beta}\delta^{\nu\alpha}) + \delta^{\alpha\beta}\del^\mu
\del^\nu - \right. \nonumber \\
&-& \left. \delta^{\alpha\nu} \del^\mu\del^\beta
-  \delta^{\mu\beta} \del^\alpha\del^\nu + \delta^{\mu\nu} \del
^\alpha\del^\beta \right] F(z-z') \A^\beta_\nu (z';x,y) \: .
\label{4.6}
\ea

\noindent
In order to write the last term between brackets in the above
expression, we used the fact that $\vec \nabla \cdot \overline
\omega =0$. In Appendix B it is shown how to integrate by parts
the derivatives $\del^\alpha$ and $\del^\beta$ in order to make
them to act on the $\overline\omega$'s of the external field $\A_\mu$.

The expression between brackets in (4.6) can be written as

\be
P^{\mu\nu\alpha\beta} = \epsilon^{\mu\alpha\rho\sigma}
\del_\sigma \epsilon^{\nu\beta\rho\lambda} \del'_\lambda \: .
\label{4.7}
\ee

\noindent
Inserting this in (4.6), integrating by parts $\del_\sigma$
and $\del'_\lambda$ (see Appendix B) and eliminating the
$\delta$-functions appearing in the external fields,
Eq. (\ref{external field}), we obtain

\ba
\Lambda^{\rm asy}(x-y) &=& -\fr{M^2}{2 g^2} \sum_{i,j=1}^2 \lambda_i
\lambda_j \int_{V_{x_i}} d^3 \xi_\mu \epsilon^{\mu \alpha \rho \sigma}
\del_{\sigma}^{(\xi)} \om^\alpha (\vec\xi -\vec x_i)
\int_{V_{x_j}} d^3 \eta_\nu \epsilon^{\nu \beta \gamma \lambda}
\del_{\lambda}^{(\eta)} \om^{\beta} (\vec\eta -\vec x_j) \times \nonumber \\
& \times & \left[ \delta^{\rho \gamma} \fr{1}{(2\pi)^2} (-\nabla^2_{(\xi)})
 \int _0 ^\infty dk \fr {e^{-\sqrt{k^2+M^2}
|x^4 -y^4|}}{\sqrt{k^2 +M^2}} \fr{\sin k |\vec \xi - \vec \eta|
}{k |\vec\xi - \vec \eta|} \right] \: ,
\label{Lambdatree2}
\ea

\noindent
where $x_1\equiv x$, $x_2\equiv y$, $\lambda_1 \equiv +1$,
$\lambda_2\equiv -1$.
In the last expression, we also made the angular and $k^4$
integrations in $F(\xi -\eta)$, Eq. (4.5).
Now, from our choice of the hypersurfaces
$V_{x_i}$, the indexes $\mu$ and $\nu$ in (\ref{Lambdatree2})
are temporal ($\mu = \nu =0$) and then we get

\ba
\Lambda^{\rm asy}(x-y) &=& -\fr{M^2}{2 g^2} \sum_{i,j=1}^2 \lambda_i
\lambda_j \int _{V_{x_i}} d^3\xi \: [\vec{\nabla}_{(\xi)}
\times \overline\omega]^l
\int_{V_{x_j}}
 d^3 \eta \: [\vec{\nabla}_{(\eta)}
\times \overline\omega]^m \times \nonumber \\
& \times & \left[\delta^{lm}
\fr{1}{(2\pi)^2} (-\nabla^2_{(\xi)})
 \int _0 ^\infty dk \fr {e^{-\sqrt{k^2+M^2}
|x^4 -y^4|}}{\sqrt{k^2 +M^2}} \fr{\sin k |\vec \xi - \vec \eta|
}{k |\vec\xi - \vec \eta|} \right] \: .
\label{4.9}
\ea

Let us now make use of the following identity

\be
\delta^{lm} \fr{\sin k |\vec x|}{|\vec x|} =
\del^l \del^m \left(\fr{1- \cos k |\vec x|}{k}\right) +
\fr{x^l x^m}{|\vec x|^2} \left(\fr{\sin k |\vec x|}{|\vec x|}
- k \cos k |\vec x| \right) \: .
\label{identity}
\ee

\noindent
Inserting this identity in the last term of (4.9), we get two
terms: $\Lambda_1$ and $\Lambda_2$ :

\ba
\lefteqn{\Lambda_{1} = -\fr{M^2}{2 g^2} \sum_{i,j=1}^2 \lambda_i
\lambda_j \int _{V_{x_i}} d^3\xi \: [\vec{\nabla}_{(\xi)}
\times \overline\omega]^l
\int_{V_{x_j}}
 d^3 \eta \: [\vec{\nabla}_{(\eta)}
\times \overline\omega]^m \: \: \times} \nonumber \\
& & \left[\fr{1}{(2\pi)^2} (-\nabla^2_{(\xi)})
 \int _0 ^\infty dk \fr {e^{-\sqrt{k^2+M^2}
|x^4 -y^4|}}{\sqrt{k^2 +M^2}} \del_{(\xi)}^l \del_{(\eta)}^m
\left( \frac{1 - \cos
k | \vec{\xi} - \vec{\eta}|}{k^2 | \vec{\xi} - \vec{\eta}|}\right) |\vec{\xi} -
\vec{\eta}| \right]
\label{Lambda1}
\ea

\noindent
and

\ba
\Lambda_{2} &=& -\fr{M^2}{2 g^2} \sum_{i,j=1}^2 \lambda_i
\lambda_j \int _{V_{x_i}} d^3\xi \: [\vec{\nabla}_{(\xi)} \times
\overline\omega]^l
\int_{V_{x_j}}
 d^3 \eta \: [\vec{\nabla}_{(\eta)} \times \overline\omega]^m
\times \nonumber \\
& \times & \left[\fr{1}{(2\pi)^2} (-\nabla^2_{(\xi)})
 \int _0 ^\infty dk \fr {e^{-\sqrt{k^2+M^2}
|x^4 -y^4|}}{\sqrt{k^2 +M^2}} \frac{(\vec{\xi} - \vec{\eta})^l
(\vec{\xi} - \vec{\eta})^m}{k |\vec{\xi} - \vec{\eta}|^2}
\left( \frac{\sin k |\vec{\xi} - \vec{\eta}|}{|\vec{\xi} - \vec{\eta}|} -
\right. \right. \nonumber \\
&-& \left. \left. k \cos k |\vec{\xi} - \vec{\eta}| \right) \right] \: .
\label{Lambda2}
\ea

In the next two subsections,
we are going to evaluate the contributions of each of them,
respectively, to the long distance behavior of the magnetic
monopole correlation function and mass.

\subsection{The $\Lambda_1$ term}

\spa
Let us consider here the contribution of
(\ref{Lambda1}) to (4.9). The derivatives $\del^l_{\xi}\del^m_\xi =
-\del^l_\xi \del^m_\eta$ can be made total derivatives, because
of they are contracted to rotationals. Then, using Gauss theorem
we can write

\ba
\Lambda_1 &=& \lim_{\rho \rightarrow \infty}
\fr{M^2}{2 g^2} \sum_{i\neq j;i,j =1}^2 \lambda_i \lambda_j
\oint _{S_i} d^2\xi_l \nabla_{(\xi)}^k [\vec{\nabla}_{(\xi)} \times
\overline\omega(\xi -x) ]^l \oint _{S_j} d^2\eta_m \nabla_{(\eta)}^k
[\vec{\nabla}_{(\eta)} \times \overline\omega(\eta -y)]^m \times \nonumber \\
& \times & \fr{1}{(2\pi)^2} \int_0^\infty dk \fr{1}{\sqrt{k^2 +M^2}}
\left(\fr{1- \cos k |\vec x-\vec y|}{k^2 |\vec x-\vec y|} \right)
|\vec x-\vec y | \: .
\label{4.13}
\ea

\noindent
In order to get (4.13), we integrated by parts the remaining
derivatives in (4.9). We can show that the extra terms vanish in the
limit $\rho \rightarrow 0$ (see Appendix D). We also considered
$x^4 = y^4$ which correspond to taking the correlation function at
equal real times when we make the analytic continuation back to
Minkowski space. Observe that we already made $\vec \xi = \vec x$
and $\vec \eta =\vec y$ in the last part of the integrand.
This is true in the local limit $\rho \rightarrow 0$. The
additional terms in the Taylor expansion around $\vec x$ and
$\vec y$ vanish in this limit. We have also dropped from (4.13)
the unphysical self-interaction terms with $i=j$. These can
be absorbed by a multiplicative renormalization of $\mu$.

Remember (4.13) is actually valid only in the large
distance limit
($|\vec x-\vec y| \rightarrow \infty$) since we
are considering only the two-leg graphs contribution.
Using the fact that

\be
\lim_{|\vec x| \rightarrow \infty}\left[\fr{1- \cos k |\vec x|}
{k^2 |\vec x|}\right] = \pi \delta(k) \: ,
\label{delta}
\ee

\noindent
we can easily perform the k-integral in (4.13). The surface
integrals in (4.13) are evaluated in Appendix C. Combining the two
results we get

\be
\Lambda_1 = - \fr{\pi^5}{32} \fr{M}{g^2} |\vec x -\vec y| \: .
\label{4.15}
\ee

\subsection{The $\Lambda_2$ term}

\spa
Let us consider now the contribution of
(\ref{Lambda2}) to (4.9). Using the identity

\be
\fr{\sin k |\vec x|}{|\vec x|} =\fr{1}{ 2} \nabla^2
\left[\fr{ 1- \cos k|\vec x|}{k} \right] - \fr{1}{2} k \cos k |\vec x| \: ,
\label{4.16}
\ee

\noindent
which can itself be obtained from (4.10), we may write

\ba
\Lambda_2 &=& -\fr{M^2}{2 g^2} \sum_{i\neq j;i,j = 1}^2
\lambda_i \lambda_j \int_{V_{x_i}} d^3 \xi \nabla_{(\xi)}^k
[\vec{\nabla}_{(\xi)} \times \overline\omega(\xi - x)]^l
\int _{V_{x_j}} d^3 \eta \nabla_{(\eta)} ^k [\vec{\nabla}_{(\eta)} \times
\overline\omega(\eta - y)]^m \times \nonumber \\
& \times &
\fr{(\xi -\eta)^l (\xi -\eta)^m}{|\vec{\xi} -\vec{\eta}|^2}
\left\{ \fr{1}{(2\pi)^2} \int_0^\infty dk \fr{1}{\sqrt{k^2+M^2}}
\left[ \fr{1}{2} \nabla_{(\xi)}^2 \left( \fr{1- \cos k|\vec \xi -\vec \eta|}
{k^2}\right) + \right. \right. \nonumber \\
&-& \left. \left.\fr{3}{2} \cos k |\vec \xi -\vec \eta|\right] \right\} \: .
\label{4.17}
\ea

Let us observe now that
since  $\vec \xi =\vec x +\vec r$ and $\vec \eta =\vec y +\vec r \; '$,
where $\vec r $ and $\vec r\;'$ are the integration variables in
$d^3\xi$ and $d^3 \eta$, we have $|\vec \xi - \vec \eta |
\tende |\vec x -\vec y|$. As a consequence, we conclude that
we can already take the limit in which the argument is large
in the term between brackets in (4.17). The last term between
brackets will vanish because of the Riemann-Lebesgue lemma. The
first one will not because of the singularity at $k =0$.
In Appendix D we show that (4.17) can be put in the
form

\ba
\Lambda_2 &=& -\fr{M^2}{2 g^2} \sum_{i\neq j;i,j = 1}^2
\lambda_i \lambda_j \oint_{S_i} d^2 \xi^l \nabla_{(\xi)}^k
[\vec{\nabla}_{(\xi)} \times \overline\omega(\xi - x)]^{j=3}
\oint _{S_j} d^2 \eta^l \nabla_{(\eta)} ^k [\vec{\nabla}_{(\eta)} \times
\overline\omega(\eta - y)]^{j=3}
\times \nonumber \\
& \times &
\fr{1}{2 (2\pi)^2} \int_0^\infty dk \fr{1}{\sqrt{k^2+M^2}}
 \left[ \fr{1- \cos k|\vec x -\vec y|}{k^2 |\vec x- \vec y |} \right]
|\vec x-\vec y| \: .
\label{4.18}
\ea

\noindent
The k-integral can be made as before. The surface integrals are
evaluated in Appendix C. The result for $\Lambda_2$ is the
same as for $\Lambda_1$, namely,

\be
\Lambda_2 = - \fr{\pi^5}{32} \fr{M}{g^2} |\vec x -\vec y| \: .
\label{4.19}
\ee

\subsection{The 0-loop result}

\spa
Collecting the contributions from $\Lambda_1$ and
$\Lambda_2$ to the large distance behavior of the
magnetic monopole two-point correlation function, we find

\be
\langle \mu(x) \mu^\dagger(y)\rangle \tende \exp \left\{ - \fr{\pi^5}{16}
\fr{M}{g^2} |\vec x-\vec y| \right\} \: .
\label{4.20}
\ee

\noindent
{}From this result, we can infer the value of the mass of the
quantum monopole at the tree-level:

\be
M_{\rm mon}^{(0)} =( \fr{\pi^4}{64}) \fr{4\pi M}{g^2} =1.522
\: \fr{4 \pi M}{g^2} \: .
\label{treelevel result}
\ee

\noindent
We see that this result is in agreement with the classical
mass of the monopole which is found to be in the range
 (1 $\leftrightarrow$ 1.787) $\fr{4 \pi M}{g^2} $ \cite{prasad,zachos}.

\section{ Conclusion }

\spa
The method of quantization of magnetic monopoles, based on
 the order disorder duality which exists between the monopole
operator and the lagrangian fields, proves to be very convenient
because of the fact that the evaluation of monopole operator
correlation functions reduces to a standard computation of quantum
field theory in the presence of an external field.

Our zero-loop result for the monopole mass falls in the range
of values which are
 obtained for the classical mass \cite{zachos}. Note however that
even at the level of our  zero loop  computation we are effectively
taking into account nontrivial quantum corrections to the monopole
correlation function.  This follows from the very fact that
we are describing the monopole excitations by means of a
fully quantized operator.

It would be very interesting to perform the same calculations
for the case of a grand-unified model, like $SU(5)$, for instance.
Also, the introduction of finite temperature effects would allow
us to study the temperature
dependence of the magnetic monopole mass,
 maybe with important cosmological consequences.

It also would be interesting to verify how the monopole
correlation functions deviate from the asymptotic large
distance regime. In order to do that one should weigh
the importance of the graphs with more than two legs which
were not considered here. It would be extremely interesting
then to study the possible effects of quantum corrections
in processes like the monopole catalysis of baryon decay
or the monopole production in spontaneous symmetry
breaking phase transitions in the early universe.

\vspace{1.5cm}

\leftline{\Large\bf Acknowledgements}

\bigskip

E.C.M. would like to thank the Physics Department of
Princeton University and especially C.Callan and D. Gross
for the kind hospitality. R.O.R. thanks M. Gleiser for useful
conversations. The authors also thank the
 Conselho Nacional de Desenvolvimento
Cient\'{\i}fico e Tecnol\'{o}gico - CNPq (Brazil) for
financial support.

\newpage

\appendix

\renewcommand{\thesection}{\Alph{section})}
\renewcommand{\theequation}{\Alph{section}.\arabic{equation}}
\setcounter{equation}{0}

\section{Two-Legs Graphs}

\spa Let us show here that graphs with two legs are the only ones
which contribute to the monopole correlation function in the long
distance regime.
Any two-legs graph can be written in a form as in (4.6):

\be
\Lambda^{(2)}= \fr{1}{g^2} \int d^3\xi_\mu \int d^3\eta_\nu \:
\om^\alpha(\xi)
\overline\omega^\beta(\eta)\left[ \epsilon^{\mu
\alpha\sigma\rho} \del_\sigma^{(\xi)} \epsilon^{\nu\beta
\lambda\gamma} \del_\lambda^{(\eta)}\right]
 \left(\vec{\nabla}_{(\xi)} \cdot \vec{\nabla}_{(\eta)}\right)
\delta ^{\gamma\rho}
F(\xi - \eta)  \: .
\label{A.1}
\ee

\noindent
Observe that since $\Lambda^{(2)}$ is dimensionless, $F(\xi - \eta)$
must have dimension of $(mass)^2$. Writing (for $\xi^4 = \eta^4$)

\be
\delta^{\gamma\rho} F(\xi -\eta) = \delta^{\gamma\rho}
\int \frac{d^3k}{(2 \pi)^4} \fr{e^{i \vec k \cdot (\vec \xi -\eta)}}{
|\vec k|^2} C^{(2)}(\vec k ;M) \:,
\label{A.2}
\ee

\noindent
we see that $C^{(2)}$ has dimension of  $mass$.

In the large distance regime, we have seen in Sec. 4 that
we can write (A.2) in the form

\be
\delta^{\gamma\rho}F(\xi - \eta) \tende [\del\del]^{\gamma\rho}
C^{(2)}(0;M) f^{(2)}(|\vec x -\vec y|) \: ,
\label{A.3}
\ee

\noindent
where $[\del\del]^{\gamma\rho}$ is either $\nabla^\gamma\nabla^\rho$
or $ \nabla^2 \hat x^\gamma \hat x^\rho$, for $\Lambda_1$ and
$\Lambda_2$, respectively. This structure remains valid for every
two-legs graph  at all orders. Only $C^{(2)}(\vec k;M)$ changes. At
0-loop, we saw that

\be
C^{(2)}(\vec k; M)_0=
\fr{M^2}{\sqrt{|\vec k|^2 + M^2}} \: .
\label{A.4}
\ee

According to what we  have seen in  Sec. 4 the  long distance behavior of
$\Lambda^{(2)}$ must be given by

\be
\Lambda^{(2)} \sim C^{(2)}(0;M) f(|\vec x- \vec y|) \: .
\label{A.5}
\ee

\noindent
Since $C^{(2)}(0;M)$ has dimension of $mass$ ($C^{(2)}(0;M)_0 =M$)
 it follows that for any two-legs
graph $f(|\vec x- \vec y|) \sim |\vec x-\vec y|$ at large distances.
This behavior leads to the exponential decay of the monopole
correlation function.

Let us consider now a generic 4-legs graph. This has the form

\ba
\Lambda^{(4)} &=& \fr{1}{g^4} \int \prod_{i=1}^4 d^3 \xi_i^{\mu_i}\
\overline\omega^{\alpha_{i}} (\xi_i) \sum_{combinations}
\left\{ \left[\epsilon^{\mu_1\alpha_1\sigma_1\rho_1} \del_{\sigma_1}^{(\xi_1)}
 \right]\ldots  \left[\epsilon^{\mu_4\alpha_4\sigma_4\rho_4}
\del_{\sigma_4}^{(\xi_4)} \right] \times \right. \nonumber \\
& \times & \left. (\nabla_{\xi_1}\cdot\nabla_{\xi_2})
(\nabla_{\xi_3}\cdot \nabla_{\xi_4})
 \delta^{\rho_1\rho_2} \delta^{\rho_3 \rho_4} \right\} F(\xi_1,\ldots ,\xi_4)
\: ,
\label{A.6}
\ea

\noindent
where $F(\xi_1, \ldots ,\xi_4)$, the analog of (A.2), is given by

\be
F(\xi_1,\ldots ,\xi_4)= \int \fr{d^3 k_1}{(2\pi)^4}\ldots
\fr{d^3k_3}{(2\pi)^4}
\fr{e^{i\vec k_1 \cdot( \vec \xi_1 -\vec \xi_4)}
e^{i\vec k_2 \cdot (\vec \xi_2 -\xi_4)} e^{i\vec k_3\cdot(\xi_3 -
\xi_4)}}{|\vec k_3|^2 (-\vec k_1\cdot \vec k_2)} C^{(4)}(\vec k_i;M) \: .
\label{A.7}
\ee

\noindent
Observe that $F(\xi_1,\ldots ,\xi_4)$ has dimension of $( mass)^4$ and
$C^{(4)}(\vec k_i;M)$  of $( mass)^{-1}$.

As in (4.5), for large distances, we can write

\be
\delta^{\rho_1\rho_2} \delta^{\rho_3\rho_4} F(\xi_1,\ldots ,\xi_4)=
[\del\del\del\del]^{\rho_1\rho_2\rho_3\rho_4}C^{(4)}(0;M)
f^{(4)}(|\vec x-\vec y|) \: ,
\label{A.8}
\ee

\noindent
where $[\del\del\del\del]^{\rho_1\rho_2\rho_3\rho_4}$ is the
obvious generalization of $[\del\del]^{\rho_1\rho_2}$
containing four derivatives. Following the same procedure as in the
case of the two-legs graphs we conclude that the large distance behavior
of a four-legs graph is given by

\be
\Lambda^{(4)}(x-y) \tende C^{(4)}(0;M) f^{(4)}(|\vec x-\vec y|) \: .
\label{A.9}
\ee

\noindent
Since $C^{(4)}$ has dimension of $(mass)^{-1}$ we conclude that
$f^{(4)} \tende |\vec x-\vec y|^{-1}$ and therefore
at large distances,

\be
\Lambda^{(4)}(x-y) \sim \fr{1}{|\vec x-\vec y|} \: .
\label{A.10}
\ee

We can easily perform the same analysis in the case of a
graph with 2n-legs. We are led, then, to the conclusion that
in general

\be
\Lambda^{(2n)}(x-y) \tende \left[{\cal M} |\vec x -\vec y| \right]^{3-2n}
\label{A.11}
\ee
where ${\cal M}$ has dimension of $mass$.
\noindent
As a consequence, we see that only two-legs graphs contribute
to the asymptotic large distance behavior of the monopole
correlation function.

\section{The Equation (4.6)}
\setcounter{equation}{0}

\spa Let us show here that the last term in (4.6) does indeed
vanish. Integrating by parts, we have an expression proportional
to

\be
\int d^4z d^4z' \del^\alpha \A^\alpha_\mu(z) \del'^\beta
\A_\nu^\beta (z') \delta^{\mu\nu} F(z-z') \:.
\label{B.1}
\ee

\noindent
We can now write

\ba
\lefteqn{\del^\alpha \A_\mu^\alpha(z) = -\frac{1}{g} \del_\alpha^{(z)}
\int_{V_x} d^3\xi_\mu \: \om^\alpha (\xi -x) \delta^4(z-\xi)=}
\nonumber \\
& & =
-\fr{1}{g}
\int_{V_x} d^3\xi_\mu \: \om^\alpha (\xi -x) \del_\alpha^{(z)}
\delta^4(z-\xi)  -\fr{1}{g}\hat{n}^\mu \oint_{S_x} d^2\xi_\alpha \: \om^\alpha
(\xi -x) \delta^4(z-\xi) \: ,
\label{B.2}
\ea

\noindent
where the last term comes from $S_x$,
the boundary of the volume $V_x$ ($\hat{n}$ is the unit vector
in the $d^3\xi^\mu$ direction.
Using Gauss' theorem in the last term of (B.2) we see that we
have, for an arbitrary function $J^\mu(z)$

\ba
\int d^4z  \del^\alpha \A^\alpha_\mu(z) J^\mu(z) &=&
-\fr{1}{g}
\int d^3\xi_\mu \: \om^\alpha(\xi -x) (-) \del_\alpha^{(\xi)}
J_\mu(\xi)  -\fr{1}{g}\int d^3\xi^\mu \: \del_\alpha^{(\xi)}
\left[\om^\alpha(\xi -x) J_\mu(\xi) \right] = \nonumber \\
& = & -\fr{1}{g}\int d^3\xi_\mu  \left[\del_\alpha^{(\xi)} \:
 \om^\alpha(\xi -x) \right]
J_\mu(\xi)
\: .
\label{B.3}
\ea

\noindent
This expression vanishes with the divergence of $\om^\alpha$.

\section{Evaluation of the Surface Integrals}
\label{angular integrations}

\setcounter{equation}{0}

\spa
Let us compute here the surface integrals appearing in
(\ref{4.13}) and (\ref{4.18}). Let us call  $I_1$ the
term involving the surface integrals appearing in (\ref{4.13}):

\begin{equation}
I_1 = \oint_{S_x} d^2 \xi^l  \nabla_{(\xi)}^k (\rot(\vec{\xi} - \vec{x}))^l
\oint_{S_y} d^2 \eta^m \nabla_{(\eta)}^k (\rot(\vec{\eta}- \vec{y}))^m \:.
\label{Bsurface1}
\end{equation}

\noindent
In spherical coordinates we have that $\om = (- \theta \sin \varphi,
\theta \cos \varphi, 0) = \theta \hat{\varphi}$. The rotational of
$\om$ is given by

\begin{equation}
\rot = \hat{r}\: \frac{1 + \theta \cot \theta}{r} - \hat{\theta}\:
 \frac{\theta}{r}
\: .
\label{Brotomega}
\end{equation}

\noindent
In (\ref{Bsurface1}), $d^2 \xi^l$ and $d^2 \eta^m$ are the elements of area
of the spherical surfaces $S_x$ and $S_y$, respectively, expressed as
$d^2 \xi^l \to r^2 d \varphi d \theta \sin \theta \: \hat{r}$ and
$d^2 \eta^m \to {r'}^2 d \varphi' d \theta' \sin \theta' \: \hat{r}'$.
{}From the integration by parts done in (\ref{4.13}), we see
that it is the radial term of $\rot$ that is going to contribute in
(\ref{Bsurface1}). We have then that

\begin{equation}
\hat{r}^l \nabla^k (\rot)_{\rm radial}^l = - \frac{\hat{r}^k}{r^2}
(1 + \theta \cot \theta) + \frac{\hat{\theta}^k}{r^2}
\left( \cot \theta - \frac{\theta}{\sin^2 \theta} \right) \: .
\label{Bvectork}
\end{equation}

\noindent
Therefore, we get for the surface integral

\begin{equation}
\oint_S d S \: \hat{r}^l \nabla^k (\rot)_{\rm radial}^l =
\int_0^{2 \pi} d \varphi \int_0^\pi d \theta \sin \theta \left[
- \hat{r}^k ( 1 + \theta \cot \theta) + \hat{\theta}^k \left(
\cot \theta - \frac{\theta}{\sin^2 \theta} \right) \right] \:.
\label{Bintegralvectork}
\end{equation}

\noindent
Substituting the vectors $\: \hat{r} = (\sin \theta \cos \varphi,
\sin \theta \sin \varphi, \cos \theta)$ and $\: \hat{\theta} =
(\cos \theta \cos \varphi, \cos \theta \sin \varphi, - \sin \theta)$
in (\ref{Bintegralvectork}) and performing the angular integrations,
one obtains

\begin{equation}
\oint_S d S \: \hat{r}^l \nabla^k (\rot)_{\rm radial}^l = \frac{\pi^3}{2}
(0, 0, 1)^k \: .
\label{Bintegralvectork2}
\end{equation}

\noindent
Thus, we get for $I_1$, Eq. (\ref{Bsurface1}), the result

\begin{equation}
I_1 = \frac{\pi^6}{4} \:.
\label{BI1}
\end{equation}

Let us call  $I_2$ the term involving the surface integrals in
(\ref{4.18}):

\begin{equation}
I_2 = \oint_{S_x} d^2 \xi^l \nabla_{(\xi)}^k (\rot(\vec{\xi}))^{j=3}
\oint_{S_y} d^2 \eta^l \nabla_{(\eta)}^k (\rot(\vec{\eta}))^{j=3} \:,
\label{Bsurface2}
\end{equation}

\noindent
where, we used the fact that $|\vec \xi -\vec \eta| \tende
|\vec x-\vec y| $ and, without loss of generality, chose
$(\vec x -\vec y)/|\vec x- \vec y|$ in the 3-direction.
 From (\ref{Brotomega}), $(\rot)^{j=3}$ is given by

\begin{equation}
(\rot)^{j=3} = \frac{\theta + \sin \theta \cos \theta}{r \sin \theta} \:.
\label{Brotomega3}
\end{equation}

\noindent
{}From (\ref{Brotomega3}) we have that

\begin{equation}
\nabla^k (\rot)^{j=3} = - \frac{\hat{r}^k}{r^2} \: \frac{\theta + \sin \theta
\cos \theta}{\sin \theta} + \frac{\hat{\theta}^k}{r^2} \:
\frac{\sin \theta \cos^2 \theta - \theta \cos \theta}{\sin^2 \theta} \: .
\label{Bvectorik}
\end{equation}

\noindent
{}From (\ref{Bsurface2}) and (\ref{Bvectorik}) we get that

\begin{equation}
\oint_S d S \: \hat{r}^l \nabla^k (\rot)^{j=3} =
\int_0^{2 \pi} d \varphi \int_0^\pi d \theta \left[ - \hat{r}^l \hat{r}^k
\: (\theta + \sin \theta \cos \theta) + \hat{r}^l \hat{\theta}^k
\: \frac{\sin \theta \cos^2 \theta - \theta \cos \theta}{\sin \theta}
\right] \:.
\label{Bintegralvectorik}
\end{equation}

\noindent
Substituting the vectors $\hat{r}$ and $\hat{\theta}$ in
(\ref{Bintegralvectorik}) and performing the angular integrations, we
obtain the double-vector:

\begin{equation}
\oint_S d S \: \hat{r}^l \nabla^k (\rot)^{j=3} =
- \frac{\pi^3}{2} \left( (1,0,0)^k, (0,1,0)^k, (0,0,0)^k \right)^l \: .
\label{Bdoublevector}
\end{equation}

\noindent
$I_2$, Eq. (\ref{Bsurface2}), is given by taking the scalar product of
(\ref{Bdoublevector}) with itself:

\begin{equation}
I_2 = \frac{\pi^6}{2} \: .
\label{BI2}
\end{equation}

\section{The Equation (4.18)}
\label{extra terms}

\setcounter{equation}{0}

 \spa Let demostrate here that Eq. (4.17) can be put in the form
(4.18). Writing \break $\nabla_{(\xi)}^2 = - \vec{\nabla}_{(\xi)} \cdot
\vec{\nabla}_{(\eta)}$
we see that (4.17) is of the form

\ba
\lefteqn{\int d^3\xi d^3\eta f(\xi) g(\eta) \vec{\nabla}_{(\xi)} \cdot
\vec{\nabla}_{(\eta)}
F(\xi -\eta) =} \nonumber \\
& &  = \int d^3\xi d^3\eta \: \left( \vec{\nabla}_{(\xi)} \cdot
\vec{\nabla}_{(\eta)}\right)
\left[f(\xi) g(\eta)
F(\xi -\eta)\right] + 3 \: \: {\rm additional \: terms} \: .
\label{D.1}
\ea

\noindent
The three aditional terms in the above expression always involve
an integral of the type

\be
\int d^3\eta T^{jkl}(\eta) \int d^3\xi \del^l_{(\xi)}\del^k_{(\xi)}
\left[\vec{\nabla} \times \overline\omega(\xi -x)\right]^i
\del^i_{(\xi)} |\xi -\eta|
F^j (\xi -\eta) \: ,
\label{D.2}
\ee

\noindent
where we used
the identity $ \frac{x^i}{|\vec x|} =
\del^i |\vec x|$.
Because of the presence of the rotational, we can make the
$\del^i$ derivative total and apply Gauss' theorem to write
the last integral as

\be
\int d^3\eta T^{jkl}(\eta) \oint d^2\xi^i \del^l_{(\xi)}\del^k_{(\xi)}
\left[\vec{\nabla} \times \overline\omega(\xi -x)\right]^i  \: \left[ |x-y|
F^j (x-y)\right] \: ,
\label{D.3}
\ee

\noindent
where we used the fact that $|\xi -\eta| \tende |x-y|$.
Making $\del^l_{(\xi)} = -\del^l_{(x)}$ when acting on
$\overline\omega(\xi -x)$ we immediately see that
the above expression is proportional to

\be
-\del^l_{(x)} \oint_{S_i} d^2\xi_i \del^k_{(\xi)}
\left[\vec{\nabla} \times \overline\omega(\xi -x) \right]^i \: .
\label{D.4}
\ee

\noindent
In Appendix C, Eq. (C.5), we showed that the above integral is a pure
number and therefore independent of $x$. As a consequence,
the above derivative vanishes and all the three aditional terms in
(D.1) are equal to zero. Using the theorem of Gauss in
the first term on the r.h.s. of (D.1) immediately leads to (4.18).
It is interesting to note that the vanishing of the three
terms which are not total derivatives in (D.1) depends
crucially on the presence of the factors $\del^i |\xi -\eta|=
(\xi -\eta)^i /|\xi -\eta|$. These terms would no longer
vanish, for instance, if we tried to apply the same procedure
of this Appendix to (4.9).

\newpage

\begin{center}
{\large \bf Figure Captions}
\end{center}

\vspace{2.0cm}

\noindent
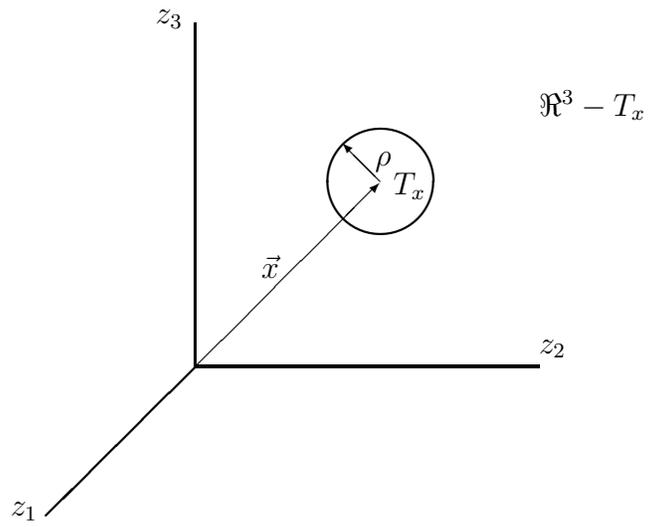
{\bf Figure 1}: Volume configuration in the definition of the external field
$\A_{\mu}^{a}(z;x)$;

\vspace{0.5cm}

\noindent
{\bf Figure 2}: Vertices involving the external field $\A_{\mu}^{a}$
(curly line) contributing at zero loop.
Vertices with dark dots represent derivative vertices
(gauge dependent). ($a$ = 1, 2);

\vspace{0.5cm}

\noindent
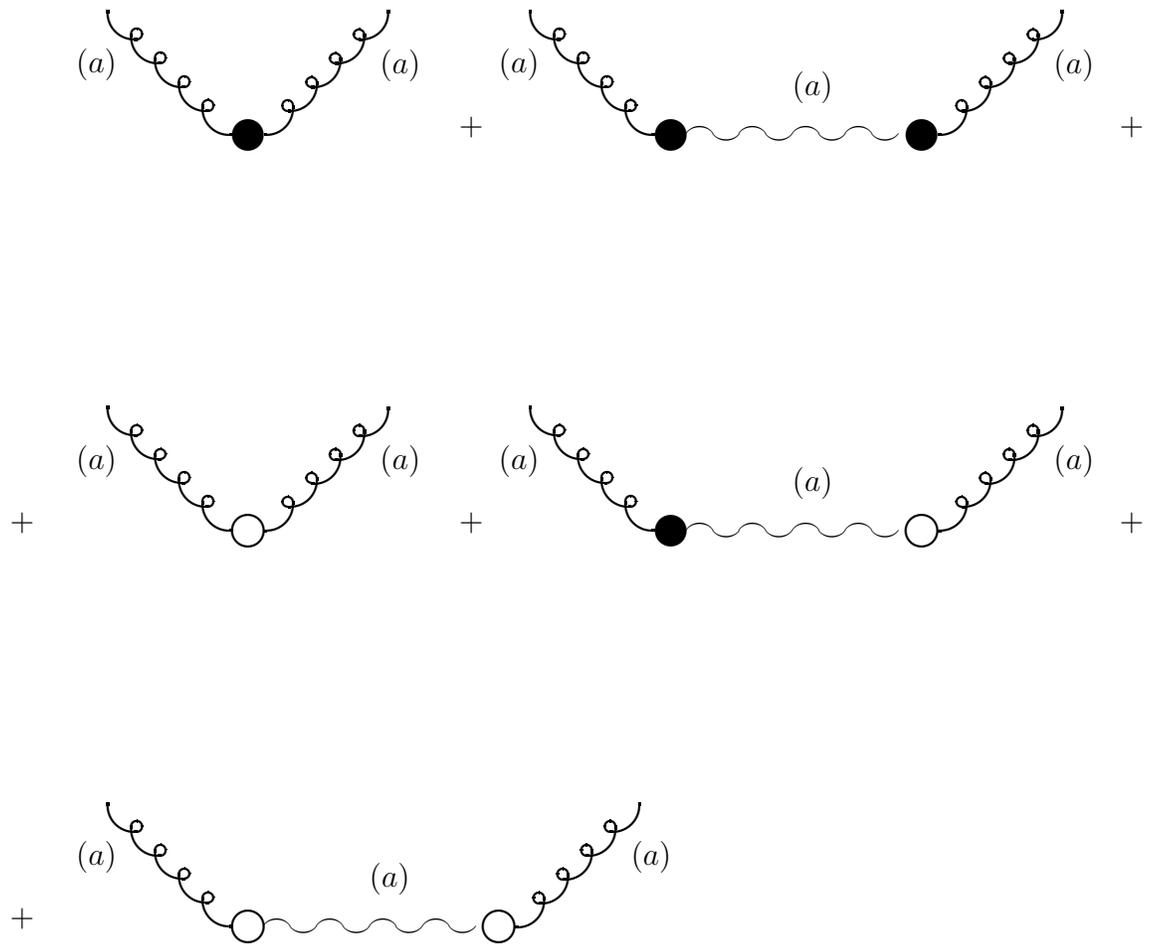
{\bf Figure 3}: Graphs contributing to the asymptotic behavior of
$\Lambda ( x - y)$,
in Eq. (4.1), at zero
loop. In the symmetric phase the contribution to $\Lambda$ is
given only by the first two graphs;


\input FEYNMAN

\newpage

\begin{picture}(20000,20000)(-16000,-10000)

\drawline\fermion[\NE\REG](-3000,-12000)[9500]
\drawarrow[\NE\ATBASE](\pbackx,\pbacky)
\put(-500,\pmidy){$\vec{x}$}
\drawline\fermion[\NW\REG](4000,-5000)[1700]
\drawarrow[\NW\ATBASE](\pbackx,\pbacky)
\put(3800,\pmidy){$\rho$}

\thicklines

\put(4000,-5000){\circle{28000}}
\put(4500,-5500){$T_{x}$}
\drawline\fermion[\N\REG](-3000,-12000)[13000]
\put(-4500,\pbacky){$z_{3}$}
\drawline\fermion[\SW\REG](\fermionfrontx,\fermionfronty)[8000]
\put(-10000,\pbacky){$z_{1}$}
\drawline\fermion[\E\REG](\particlefrontx,\particlefronty)[13000]
\put(\pbackx,-11500){$z_{2}$}
\put(10000,-2500){$\Re^{3} - T_{x}$}
\put(5000,-43000){Figure 1}
\end{picture}

\newpage

\begin{picture}(20000,20000)(-10000,-10000)
\thicklines
\bigphotons
\put(-1000,5000){\circle*{1200}}
\drawline\gluon[\NW\REG](-1600,5000)[4]
\put(-7500,\pmidy){$(a)$}
\drawline\gluon[\NE\FLIPPED](-400,5000)[4]
\put(4000,\pmidy){$(a)$}
\put(20000,5000){{\large $-\frac{\xi}{2}(\partial_{\mu}\A_{a}^{\mu})^2$}}

\put(-1000,-7000){\circle*{1200}}
\drawline\gluon[\NW\REG](-1600,-7000)[4]
\put(-7500,\pmidy){$(a)$}
\drawline\photon[\E\REG](-400,-7000)[7]
\put(\pmidx,-5500){$(a)$}
\put(20000,-7000){{\large $\xi \partial_{\mu} \A_{a}^{\mu} \partial_{\nu}
W_{a}^{\nu}$}}

\put(-1000,-19000){\circle{1200}}
\drawline\gluon[\NW\REG](-1600,-19000)[4]
\put(-7500,\pmidy){$(a)$}
\drawline\gluon[\NE\FLIPPED](-400,-19000)[4]
\put(4000,\pmidy){$(a)$}
\put(20000,-19000){{\large $-\frac{M^2}{2} (\A_{a}^{\mu})^2$}}

\put(-1000,-31000){\circle{1200}}
\drawline\gluon[\NW\REG](-1600,-31000)[4]
\put(-7500,\pmidy){$(a)$}
\drawline\photon[\E\REG](-400,-31000)[7]
\put(\pmidx,-29500){$(a)$}
\put(20000,-31000){{\large $M^2 \A_{a}^{\mu} W_{a}^{\mu}$}}

\put(10000,-43000){Figure 2}

\end{picture}

\newpage

\begin{picture}(20000,20000)(-10000,-10000)
\thicklines

\put(-1000,5000){\circle*{1200}}
\drawline\gluon[\NE\FLIPPED](-400,5000)[4]
\put(4000,\pmidy){$(a)$}
\drawline\gluon[\NW\REG](-1600,5000)[4]
\put(-7500,\pmidy){$(a)$}
\put(7000,5000){+}
\bigphotons
\put(15000,5000){\circle*{1200}}
\put(24500,5000){\circle*{1200}}
\drawline\photon[\E\REG](15600,5000)[8]
\put(\pmidx,6500){$(a)$}
\drawline\gluon[\NE\FLIPPED](25100,5000)[4]
\put(29500,\pmidy){$(a)$}
\drawline\gluon[\NW\REG](14400,5000)[4]
\put(8500,\pmidy){$(a)$}
\put(32000,5000){+}

\put(-10000,-10000){+}
\put(-1000,-10000){\circle{1200}}
\drawline\gluon[\NW\REG](-1600,-10000)[4]
\put(-7500,\pmidy){$(a)$}
\drawline\gluon[\NE\FLIPPED](-400,-10000)[4]
\put(4000,\pmidy){$(a)$}
\put(7000,-10000){+}
\put(15000,-10000){\circle*{1200}}
\put(24500,-10000){\circle{1200}}
\drawline\photon[\E\REG](15600,-10000)[8]
\put(\pmidx,-8500){$(a)$}
\drawline\gluon[\NE\FLIPPED](25100,-10000)[4]
\put(29500,\pmidy){$(a)$}
\drawline\gluon[\NW\REG](14400,-10000)[4]
\put(8500,\pmidy){$(a)$}
\put(32000,-10000){+}

\put(-10000,-25000){+}
\put(-1000,-25000){\circle{1200}}
\put(8500,-25000){\circle{1200}}
\drawline\gluon[\NW\REG](-1600,-25000)[4]
\put(-7500,\pmidy){$(a)$}
\drawline\photon[\E\REG](-400,-25000)[8]
\put(\pmidx,-23500){$(a)$}
\drawline\gluon[\NE\FLIPPED](9100,-25000)[4]
\put(13500,\pmidy){$(a)$}

\put(10000,-43000){Figure 3}
\end{picture}

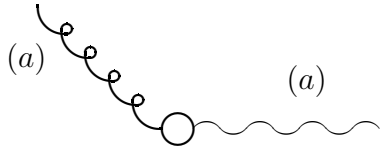
\end{document}